\documentclass[twocolumn,aps,prc,floatfix]{revtex4}
\usepackage{graphicx}
\usepackage{dcolumn}
\usepackage{bm}
\usepackage{mhchem}
\usepackage{color}
\begin{document}

\title{Double strangeness $\Xi^{-}$ production as a probe of nuclear equation of state \\at high densities}

\author{Gao-Chan Yong$^{1,2}$}
\email[]{yonggaochan@impcas.ac.cn}
\author{Zhi-Gang Xiao$^{3}$}
\author{Yuan Gao$^{4}$}
\author{Zi-Wei Lin$^{5}$}
\affiliation{$^{1}$Institute of Modern Physics, Chinese Academy of Sciences, Lanzhou 730000, China \\
$^{2}$School of Nuclear Science and Technology, University of Chinese Academy of Sciences, Beijing, 100049, China}
\affiliation{$^{3}$Department of Physics, Tsinghua University, Beijing 100084, China}
\affiliation{$^{4}$School of Information Engineering, Hangzhou Dianzi University, Hangzhou 310018,
People's Republic of China}
\affiliation{$^{5}$Department of Physics, East Carolina University, Greenville, North Carolina 27858, USA}

\begin{abstract}

Double strangeness $\Xi^{-}$ production in Au+Au collisions at 2, 4, and 6 GeV/nucleon incident beam energies is studied with the pure hadron cascade version of a multi-phase transport model.
It is found that due to larger nuclear compression, the model with the soft  equation of state (EoS) gives larger yields of both single strangeness ($K^{+}$ and $\Lambda+\Sigma^{0}$) and double strangeness $\Xi^{-}$. The sensitivity of the double strangeness $\Xi^{-}$ to the EoS is evidently larger than that of $K^{+}$ or $\Lambda+\Sigma^{0}$ since the phase-space distribution of produced $\Xi^{-}$ is more compact compared to those of the single strangeness. The larger sensitivity of the yields ratio of $\Xi^{-}$ to the EoS from heavy and light systems is kept compared to that of the single strangeness. The study of $\Xi^{-}$ production in relativistic heavy-ion collisions provides an alternative for the ongoing heavy-ion collision program at facilities worldwide for identifying the EoS at high densities, which is relevant to the investigation of the phase boundary and onset of deconfinement of dense nuclear matter.

\end{abstract}

\maketitle

\section{Introduction}
Understanding the properties of nuclear matter under conditions of extreme energy and baryon density is one of the main goals of the relativistic heavy-ion collision program \cite{pr1}. Constraining the equation of state (EoS) of dense nuclear matter has been a longstanding and shared goal of both nuclear physics and astrophysics \cite{as1,as2,as3,as4,as5,as6}. The EoS of dense nuclear matter plays a crucial role in understanding the later evolution of the high-temperature quark-gluon-plasma (QGP) created after the Big Bang and the physics associated with compact stars as well as the QCD phase diagram. Strange mesons or baryons are excellent probes for identifying the EoS, the phase boundary and onset of deconfinement \cite{kaon2006,raf82,adam20}, which are currently being carried out or planned at CBM and HADES at GSI/FAIR, BM@N and MPD at NICA, CEE at HIAF, Dipole Hadron Spectrometer and Dimuon Spectrometer at J-PARC-HI, and STAR at RHIC BES-II \cite{qm2018}.

Associations of the single strangeness kaon meson with the EoS have been extensively studied in the literature \cite{fuchs2001,kaon2006,ditoro,fuchs2006,jor1985,feng11}. The single strangeness $\Lambda$ hyperons have been measured by the E895 and FOPI Collaborations \cite{chung2001,thesis2004,liq05}, while the double strangeness $\Xi^{-}$ production in heavy-ion collisions has been studied extensively in the last twenty years \cite{exp2003,exp2009,exp2015,ko2002npa,ko2004npa,ko2004plb,chen2004plb,li2012prc,urqmd2014,urqmd2016,buu2018}. However, their connections to the dense nuclear EoS are less studied.
Due to strangeness conservation, once produced, $K^{+}$ and $K^{0}$ mesons are rarely absorbed by the surrounding nucleons in low energy nuclear collisions where
there are far fewer $K^{-}$ and $\bar{K^{0}}$ mesons. The lack of final state interactions makes kaon a penetrating probe for the EoS of dense matter produced in heavy-ion collisions. Since the kaon meson mainly comes from the compressed high-density region \cite{fuchs2001,kaon2006,ditoro} and the double strangeness $\Xi^{-}$ comes mainly from the collision of two singly strange particles \cite{li2012prc,urqmd2014,buu2018}, it is naturally to question whether the double strangeness $\Xi^{-}$ production is more sensitive to the properties of the dense matter created in heavy-ion collisions. Therefore, in this Letter we study the phase-space distribution of produced double strangeness $\Xi^{-}$ and its dependence to the EoS of the dense matter formed in heavy-ion collisions.

A multi-phase transport (AMPT) model is recently extended so that it can perform not only multi-phase transport simulations with both parton and hadron degrees of freedom \cite{AMPT2005} but also pure hadron cascade with hadronic mean-field potentials for low energy heavy-ion collisions. Since this study mainly focuses on the double strangeness $\Xi^{-}$ production in low energy nuclear collisions, we have also added more channels of $\Xi^{-}$ interactions to the AMPT model \cite{li2012prc,urqmd2014,buu2018}. Our studies show that the phase-space distribution of produced $\Xi^{-}$'s in heavy-ion collisions is more compact than the single strangeness $K^{+}$ or $\Lambda+\Sigma^{0}$, therefore the $\Xi^{-}$ production in heavy-ion collisions is more sensitive to the EoS of the dense matter than single strangeness hadrons.

\section{Brief description of the AMPT-HC model}

The AMPT model consists of four main components: a fluctuating initial condition, partonic interactions, conversion from the partonic to the hadronic matter, and hadronic interactions \cite{AMPT2005}. As a Monte Carlo parton and hadron transport model, it has been extensively
applied to high-energy heavy-ion collisions and the capability to reproduce multi experimental observables has been demonstrated. In the hadron cascade of the AMPT model, the following hadrons
are explicitly included: $\pi$, $\rho$, $\omega$, $\eta$, $K$, $K^*$,
and $\phi$ for mesons; $N$, $\Delta$, $N^*(1440)$, $N^*(1535)$, $\Lambda$,
$\Sigma$, $\Xi$, and $\Omega$ and deuteron for baryons and light nuclei including the  corresponding antiparticles. Many interactions among these hadrons and the corresponding inverse reactions are included \cite{AMPT2005,deu2009}.

In order to describe heavy ion collisions at low beam energies,
we have extended the AMPT model to a new pure hadron cascade version (AMPT-HC).
First, the Woods-Saxon nucleon density distribution and local Thomas-Fermi approximation are used to initialize the position and momentum of each nucleon in projectile and target nuclei. Then parton interactions and the conversion from the partonic matter to the hadronic matter are switched off. Finally, hadron potentials with the test-particle method are applied
to nucleons, baryonic resonances, $K$, $\Lambda$, $\Sigma$, $\Xi$ and $\Omega$ as well as their antiparticles in addition to the usual elastic and inelastic collisions. In this study, the form of kaon potential was taken from Ref.~\cite{ligq97}. For strange baryons $\Lambda$, $\Sigma$, $\Xi$, their mean-field potentials are 2/3, 2/3, and 1/3, respectively, of that of nucleon according to the quark counting rule \cite{ligq98}.

In order to describe the double strangeness $\Xi^{-}$ production in low energy nuclear collisions, besides the strangeness exchange reactions $\bar{K}$+$Y$ $\leftrightarrow$ $\pi$+$\Xi$ (Y = $\Lambda$ or $\Sigma$) that are already in the public AMPT model \cite{AMPT2005,website}, the isospin-averaged reactions $Y$+$Y$ $\leftrightarrow$ $N$+$\Xi$ are included \cite{li2012prc,urqmd2014}. In addition, we have added the production of $\Xi$ via $Y$+$N$ $\rightarrow$ $N$+$\Xi$+$K$ \cite{buu2018}. Note that we have not yet added the corresponding anti-baryon channels for these newly added interactions
to the AMPT model, where we expect their effects to be small at low energies for this study.
Also note that $\Sigma^{0}$ is very unstable and decays into $\Lambda$ and $\gamma$; therefore, we analyze $\Lambda$ and $\Sigma^{0}$ together. In the study, the hadronic soft ($k_{0}$ = 200 MeV) and stiff ($k_{0}$ = 300 MeV) EoSs as well as a first-order phase transition EoS are applied to investigate the strangeness productions.

\section{Results and Discussions}

\begin{figure}[t]
\centering
\includegraphics[height=4cm,width=0.45\textwidth]{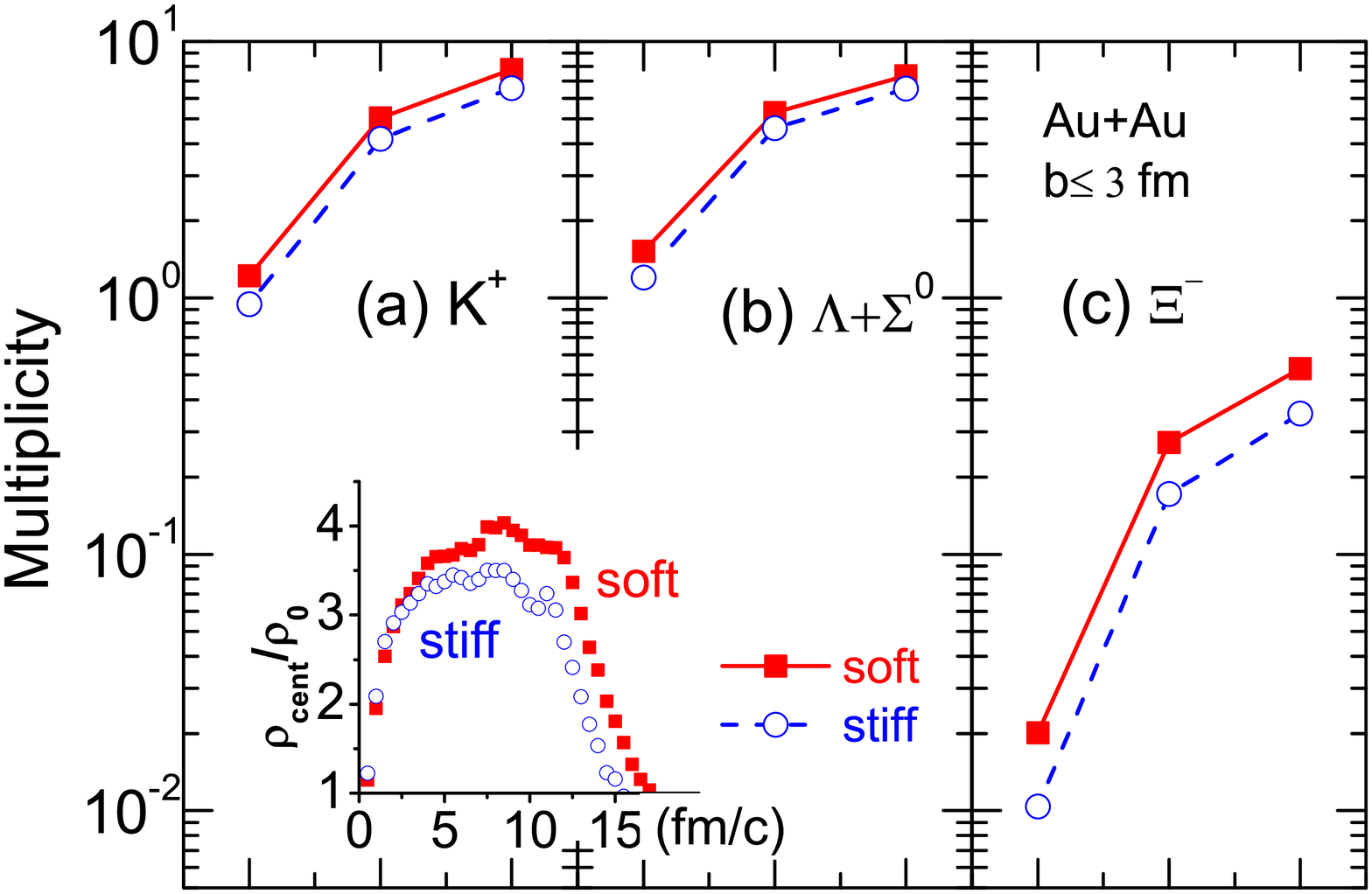}
\includegraphics[height=4cm,width=0.45\textwidth]{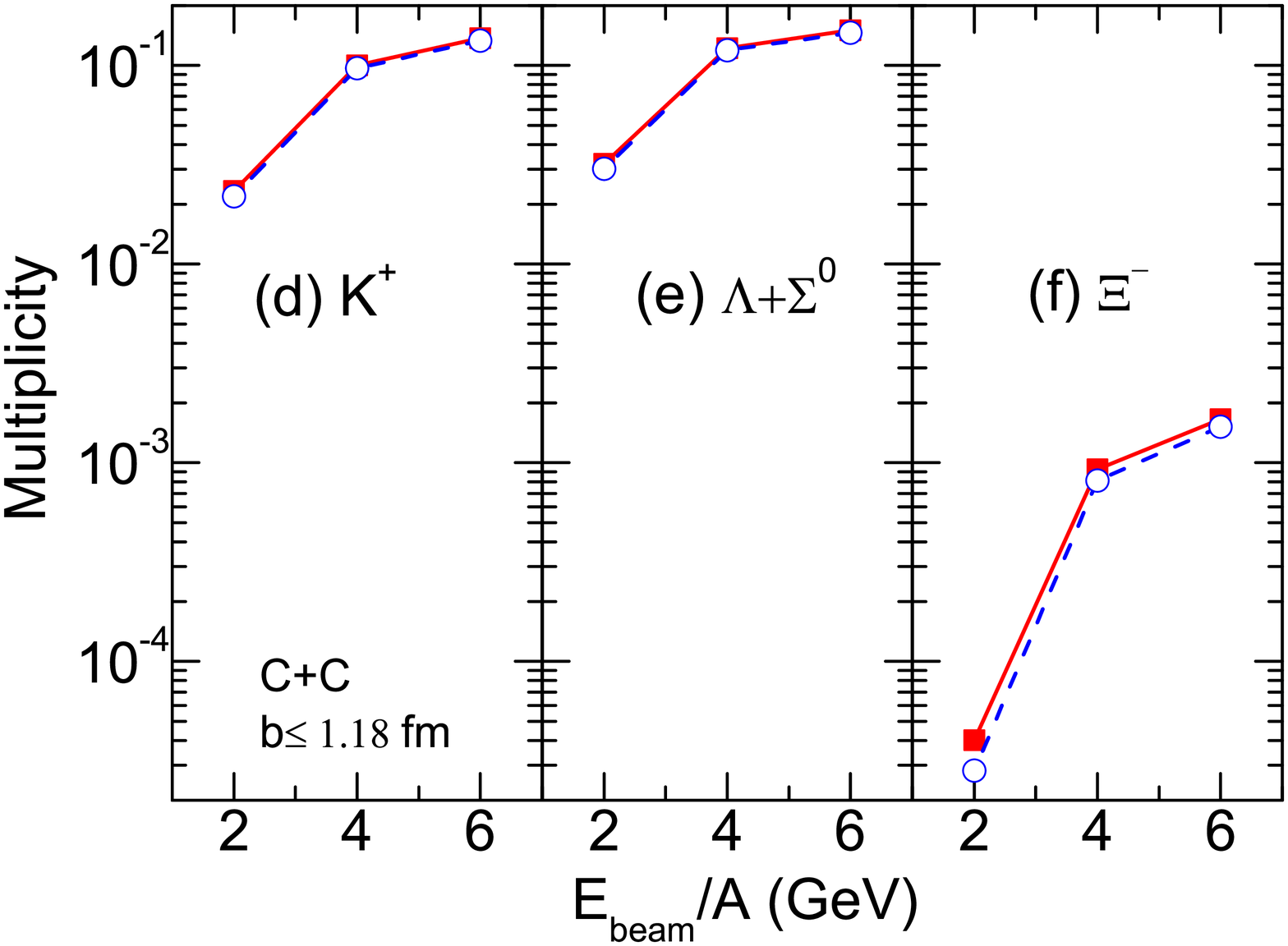}
\caption{Top: yields of $K^{+}$ (a), $\Lambda+\Sigma^{0}$ (b) and $\Xi^{-}$ (c) in central Au+Au reactions at incident beam energies of 2, 4, and 6 GeV/nucleon with soft ($k_{0}$ = 200 MeV) and stiff ($k_{0}$ = 300 MeV) EoSs; the inset shows evolutions of the central compression densities with soft and stiff EoSs at 2 GeV/nucleon. Bottom: same as top panels but for the C+C system.} \label{yield}
\end{figure}
The connections between the kaon production and the EoS have been studied in the literature \cite{fuchs2001,kaon2006,ditoro,fuchs2006,jor1985}. Here we extend our studies to double strangeness production in Au+Au and C+C collisions at 2-6 GeV/nucleon beam energies. Fig.~\ref{yield} shows the yields of $K^{+}$, $\Lambda+\Sigma^{0}$ and $\Xi^{-}$ in central Au+Au (top) and C+C (bottom) reactions at incident beam energies of 2, 4, 6 GeV/nucleon with soft ($k_{0}$ = 200 MeV) and stiff ($k_{0}$ = 300 MeV) EoSs. Since the single nucleon potential at high energies (above 1 GeV kinetic energy) and high densities (above 3 times saturation density) is still undetermined \cite{guo2021}, to make minimum assumptions we use the density-dependent form \cite{Ga01} for the single nucleon mean-field potential, i.e.,
\begin{equation}
U(\rho)=\alpha\frac{\rho}{\rho_0}+\beta \left (\frac{\rho}{\rho_0} \right)^\gamma,
\label{potential}
\end{equation}
where $\rho_0$ stands for the saturation density. We use
\begin{eqnarray}
\alpha &=& \left (-29.81 - 46.9\frac{k_{0}+44.73}{k_{0}-166.32} \right ) {\rm MeV},\nonumber \\
\beta &=& 23.45\frac{k_{0}+255.78}{k_{0}-166.32}  {~\rm MeV}, \nonumber \\
\gamma &=& \frac{k_{0}+44.73}{211.05}
\label{coeff}
\end{eqnarray}
to model the soft and stiff EoSs.
Note that the symbol $k_{0}$ in Eq.~\eqref{coeff} represents the $k_{0}$ value in MeV,
and the potential $U(\rho)$ has the same value at density $\rho_0$ but splits at higher densities.
The inset of Fig.~\ref{yield} shows the time evolution of
the compression density of the central cell ($|x|\leq0.5$ fm, $|y|\leq0.5$ fm, $|z|\leq0.5$ fm)
with soft and stiff EoSs at 2 GeV/nucleon.
It demonstrates that, due to weaker nuclear repulsion, the soft EoS causes a larger compression than the stiff EoS. From the upper panels of Fig.~\ref{yield}, it is seen that, due to larger compression, the AMPT-HC model with the soft EoS gives larger yields
of both single strangeness ($K^{+}$ and $\Lambda+\Sigma^{0}$) and double strangeness $\Xi^{-}$. Since the single strangeness is mainly produced from the collisions of two non-strange hadrons
while the double strangeness is mainly produced from the collisions of two single strangeness hadrons, the yields of the double strangeness $\Xi^{-}$ are much lower than those of the single strangeness $K^{+}$ or $\Lambda+\Sigma^{0}$. The sensitivity of $K^{+}$ to the EoS is almost the same as that of $\Lambda+\Sigma^{0}$. Comparing Fig.~\ref{yield}(c) with panels (a) and (b), one sees that the sensitivity of the double strangeness $\Xi^{-}$ to the EoS is evidently larger than that of $K^{+}$ or $\Lambda+\Sigma^{0}$. The double strangeness $\Xi^{-}$ thus can serve as a probe of the EoS of dense matter with enhanced sensitivity. The lower panels of Fig.~\ref{yield} show the effect of the EoS on the strangeness production in the C+C collision system. It is found that, due to smaller space-time volume of the dense matter, the EoS has smaller effects on these strange particle productions. Therefore, heavy systems are preferred for probing the EoS of the dense matter with strangeness, colorred which is similar with the effect of probing $Esym(\rho)$ with pion yield ratio at sub-GeV energy domain \cite{mzhang09}.

\begin{figure}[t!]
\centering
\includegraphics[width=0.48\textwidth]{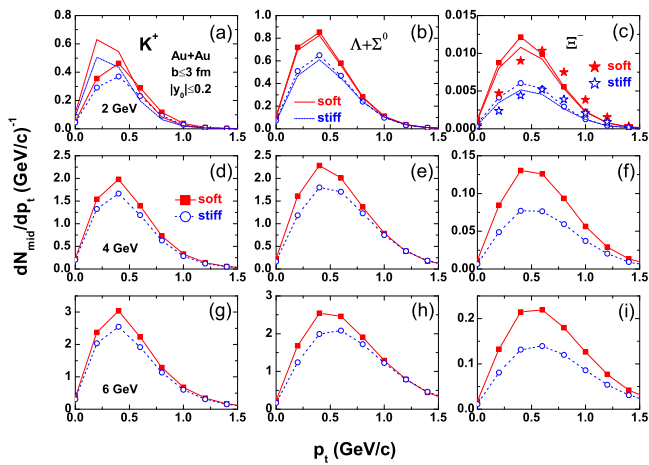}
\caption{Transverse momentum distributions of $K^{+}$ (a), $\Lambda+\Sigma^{0}$ (b) and $\Xi^{-}$ (c) around mid-rapidity in central Au+Au reactions at various incident beam energies with soft ($k_{0}$ = 200 MeV) and stiff ($k_{0}$ = 300 MeV) EoSs;
$y_{0} = y/y_{beam}$ where both $y$ and $y_{beam}$ are rapidity values in the center-of-mass frame. The solid and dished lines without symbols in panels (a), (b) and (c) show effects of mean field that couples to strange hadrons, see text for details. The filled and open stars without lines in panel (c) show the cases of 14\% decrease of $\Xi$ mass with the soft and stiff EoSs, respectively.
} \label{rapdis}
\end{figure}
To see more clearly the effects of the EoS on single or double strangeness productions, we plot in Fig.~\ref{rapdis} the transverse momentum distributions of strangeness around mid-rapidity with different EoSs. It is seen that, around the peak of the strangeness transverse momentum distribution, there is a maximum sensitivity of strangeness productions to the EoS. At 2 GeV/nucleon, for example, the maximum sensitivities to the EoS (as measured by the increase of the $p_t$-differential yield) for $K^{+}$, $\Lambda+\Sigma^{0}$ and $\Xi^{-}$ are 24\%, 32\% and 100\%, respectively. Therefore, the sensitivity of the double strangeness $\Xi^{-}$ production to the EoS is about 3 $\sim$ 4 times that of the single strangeness $\Lambda+\Sigma^{0}$ or $K^{+}$, which clearly demonstrates the advantage of using double strangeness $\Xi^{-}$ in probing the EoS of dense matter.

At 2 GeV/nucleon, we examined effects of mean field ($U_{K^{+}}, U_{\Lambda+\Sigma^{0}}, U_{\Xi^{-}}$) that couples to strange hadrons on strangeness productions by turning off $U_{K^{+}}$ and changing the mean field that couples to strange baryons into nucleonic mean field. Since at incident beam energies studied here, strange quarks are less produced, we did not consider mean field that couples to strange quarks in the present AMPT-HC model. From panels (a), (b) and (c) in Fig.~\ref{rapdis}, it is seen that the mean field that couples to strange hadrons does not evidently affect $\Lambda+\Sigma^{0}$ and $\Xi^{-}$ productions except for $K^{+}$. The larger effects of the mean field $U_{K^{+}}$ on kaon production complicate its ability to infer the EoS. Consequently, one cannot precisely probe the EoS by using kaon production if the kaon potential is not well determined. From upper panels of Fig.~\ref{rapdis}, it is also seen that the mean field that couples to strange hadrons has negligible effects on the sensitivities of the transverse momentum distributions of strangeness to the EoS. Therefore, the enhanced sensitivity of the double strangeness $\Xi^{-}$ production to the EoS is not due to its strange mean field. To check if the heavier mass of $\Xi^{-}$ compared to single strangeness is the reason of its larger sensitivity to the EoS, we show in panel (c) of Fig.~\ref{rapdis} the effects of $\Xi$ mass on the $\Xi^{-}$ production, where stars without lines show our results
after decreasing the $\Xi$ mass almost to the $\Lambda$ mass. It demonstrates that the smaller $\Xi$ mass does not affect its sensitivity to the EoS. Therefore, it seems that the enhanced sensitivity of the double strangeness $\Xi^{-}$ production to the EoS results from its production in multi-step collisions due to its double strangeness.

\begin{figure}[th]
\centering
\includegraphics[width=0.5\textwidth]{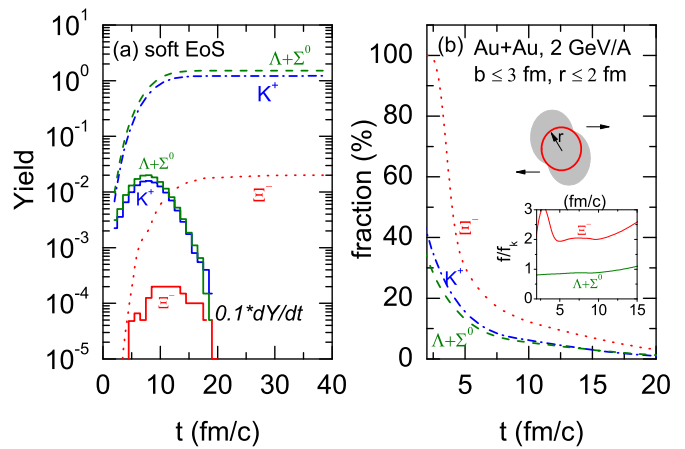}
\caption{Time evolutions of (a) strangeness production and (b) fractions of strangeness in the central cell with r $\leq$ 2 fm in central Au+Au reactions at the incident beam energy of 2 GeV/nucleon with the soft EoS (the inset shows the fractions of $\Lambda+\Sigma^{0}$ and $\Xi^{-}$ over the $K^{+}$ fraction as a function of time).
} \label{central}
\end{figure}
To see why the double strangeness $\Xi^{-}$ is more sensitive to the EoS than the single strangeness $K^{+}$ and $\Lambda+\Sigma^{0}$, we further examine in Fig.~\ref{central} the spatial and temporal distributions of strangeness productions in Au+Au collisions at the incident beam energy 2 GeV/nucleon. From the left panel of Fig.~\ref{central}, it is seen that the yields of strangeness are almost saturated at 15 fm/c when the central compression density decreases to normal nuclear density $\rho_{0}$ as shown in the inset of Fig.~\ref{yield}. The rate of the strangeness yield $dY/dt$ (solid color curves) indicates that almost all the strange particles studied here are produced at suprasaturation densities, especially at densities above $3\rho/\rho_{0}$. So these strange particles are suitable for probing the EoS at high densities. In addition, we show in the right panel of Fig.~\ref{central} the time evolution of the fractions of strangeness in the central cell (with $r \leq$ 2 fm) of the overlap volume. It is seen from the inset that the fraction of $\Xi^{-}$ in the central cell is usually more than twice the fraction of $K^{+}$ or $\Lambda+\Sigma^{0}$. Since $\Xi^{-}$ has two strangeness and cannot be easily produced from one primary collision (e.g. $NN\rightarrow \Xi K K N$ or $NN\rightarrow \Xi K Y$, these channels are not included in the present AMPT-HC model due to their negligible effects), it's dominated by multi-step collisions (first produce single strange $K$ or $Y$, then two single strange hadrons need to collide to create $\Xi$). This will naturally prefer high density region since most interactions happen there, and naively the sensitivity of $\Xi^{-}$ to the EoS is double the sensitivity of single strangeness to the EoS. In addition, this picture means that $\Xi^{-}$ will be produced later than $K$ or $Y$, which can indeed be seen in Fig.~\ref{central} (a).

\begin{figure}[t!]
\centering
\includegraphics[width=0.5\textwidth]{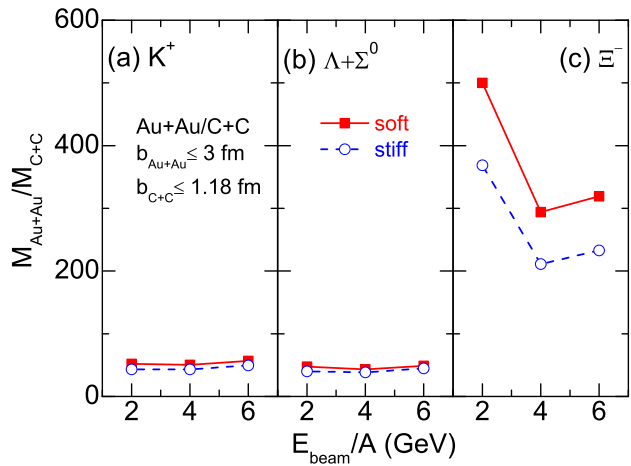}
\caption{Ratios of the yields of $K^{+}$ (a), $\Lambda+\Sigma^{0}$ (b) and $\Xi^{-}$ (c) in central Au+Au over C+C reactions at incident beam energies of 2, 4, and 6 GeV/nucleon with soft ($k_{0}$ = 200 MeV) and stiff ($k_{0}$ = 300 MeV) EoSs.
} \label{double}
\end{figure}
To reduce systematic errors and some uncertainties (especially about the cross section of $\Xi^{-}$ production around its threshold), it is meaningful to construct the ratio of the strangeness productions from heavy and light systems \cite{fuchs2001,fuchs2006}. Fig.~\ref{double} shows the ratio of strangeness productions of $K^{+}$, $\Lambda+\Sigma^{0}$ and $\Xi^{-}$ in Au+Au and C+C systems. The sensitivity of the yields ratio of $\Xi^{-}$ (35--37\% from 2 to 6 GeV/nucleon) in the two systems is much pronounced than that of the single strange yield ratios (20--14\% for $K^{+}$ and 20--11\% for $\Lambda+\Sigma^{0}$). Fig.~\ref{double} thus demonstrates that the ratio of the double strange $\Xi^{-}$ yields from heavy and light systems can be used to probe the EoS of dense matter formed in heavy-ion collisions.

\begin{figure}[th]
\centering
\includegraphics[width=0.5\textwidth]{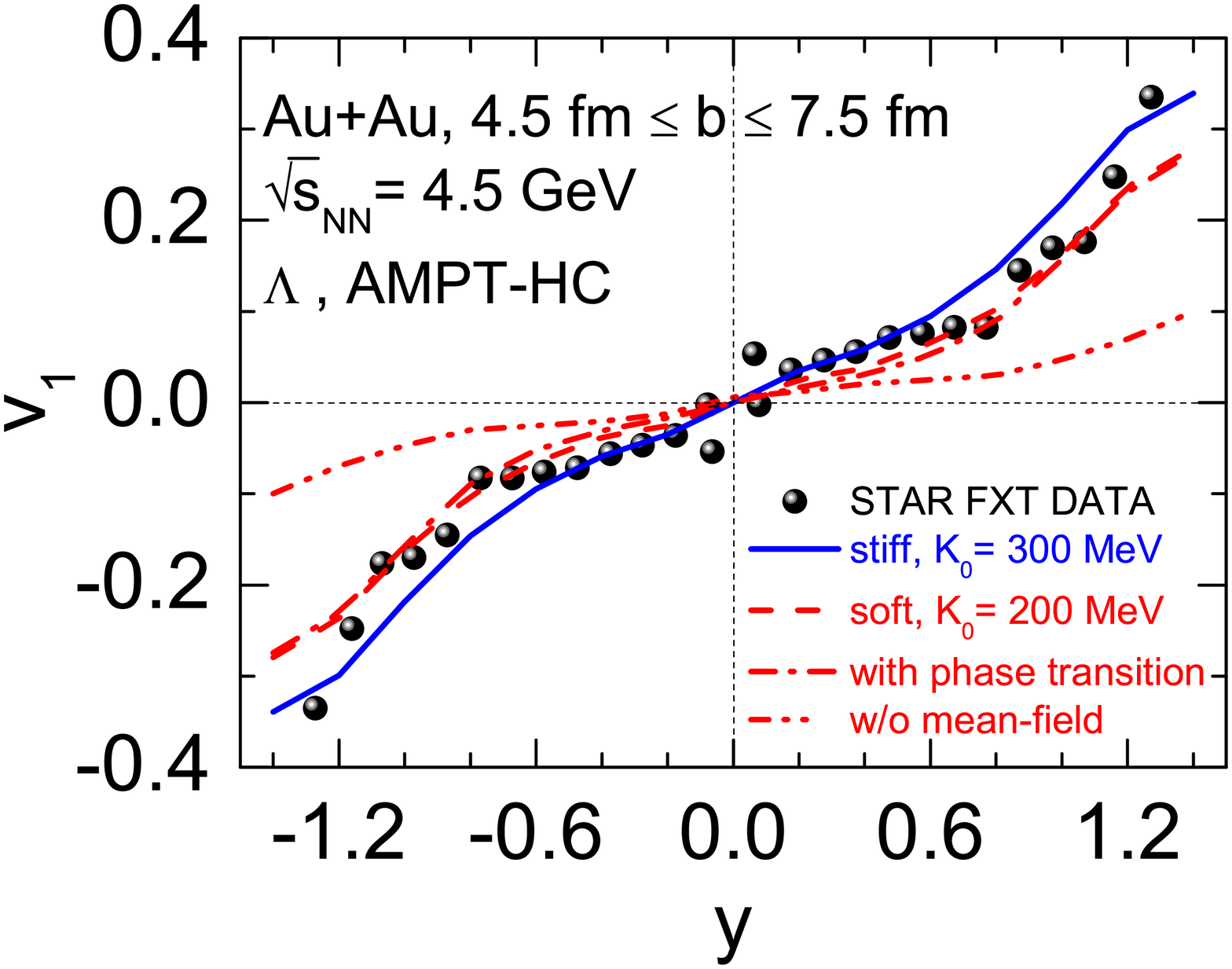}
\caption{Rapidity dependence of the directed flow $v_{1}$ of $\Lambda$ with different EoSs and different phases in comparison with the STAR FXT Au+Au data \cite{bes2021}. All the values in the negative rapidity region are reflected.} \label{flow}
\end{figure}
While studying the EoS of dense matter, it is crucial to see if the phase transition from hadronic to partonic phase occurs. To this end, a comparison is made between the directed flows $v_{1} (= \langle p_{x}/p_{t}\rangle$) for the $\Lambda$ hyperon calculated with different EoSs and  different phases and the recent STAR FXT Au+Au data. The $\Lambda$ particles analyzed here include the final state $\Sigma^{0}$ decays. From Fig.~\ref{flow}, it is seen that the data are within the range of the model results with either stiff or soft hadronic EoSs. In the mid-rapidity region, the slope of $v_{1}$ data agrees well with the AMPT-HC result with the stiff EoS, while at larger rapidities the soft EoS gives a better description of the STAR FXT Au+Au data. It is interesting to see that the AMPT-HC result with a first-order phase transition EoS \cite{guo2021}, which is given by Eq.~\eqref{potential} with $k_{0}$= 230 MeV below $2.5\rho/\rho_{0}$, an isobaric EoS at $2.5\leq\rho/\rho_{0}\leq4$ and the MIT bag model with B= $150 MeV/fm^{3}$ and $\alpha_{s}= 0.1$ above $4\rho/\rho_{0}$, is very similar to the result with the soft hadronic EoS. In addition, we see that the results without the mean field fail to describe the data, indicating the importance of the mean field in heavy-ion collisions at this energy range.
From Fig.~\ref{flow}, one cannot deduce whether a phase transition from hadronic to partonic phase occurs at this single collision energy. Data at multiple collision energies will be needed, where the beam energy scan (BES-II) program has been being carried out at RHIC-STAR \cite{qm2018}.

\section{Conclusions}

We have studied the productions of single and double strangeness particles $K^{+}$, $\Lambda+\Sigma^{0}$ and $\Xi^{-}$ in relativistic heavy-ion collisions at incident beam energies of 2, 4, and 6 GeV/nucleon covered by various new facilities around the world. It is found that all these strange particles are mainly produced from high-density region and thus carry crucial information of the dense matter. Using a new pure hadron cascade version of the AMPT model,
we show that the double strangeness $\Xi^{-}$
has a larger fraction coming from the most central cell of the collision geometry; consequently it has a significant advantage over single strangeness hadrons in probing the EoS of dense nuclear matter. To reduce systematic errors and uncertainties of the $\Xi^{-}$ production, the ratio of the $\Xi^{-}$ yields from heavy and light systems has been constructed and exhibits more pronounced sensitivity on the
EoS compared to that of single strangeness. The double strangeness $\Xi^{-}$ production in relativistic heavy-ion collisions can thus be utilized to decode the properties of nuclear matter under conditions of extreme energy and baryon densities.

\section{Acknowledgments}

This work is supported in part by the National Natural Science Foundation of China under Grant Nos. 11775275, 11890712, 11875013, and by the Ministry of Science and Technology under Grant No. 2020YFE0202001 and the National Science Foundation under Grant No. 2012947.

\end{document}